\documentstyle[aps,prl,epsf,floats]{revtex}

\begin{document}

\draft


\title{Correlated Initial Conditions in Directed Percolation}
\author{Haye Hinrichsen$^1$ and G\'eza \'Odor$^2$\\}
\address{$^1$Max-Planck-Institut f\"ur Physik komplexer Systeme,
          N\"othnitzer Stra{\ss}e 38, 01187 Dresden, Germany}
\address{$^2$ Research Institute for Technical Physics and
        Materials Science,
          P. O. Box 49, H-1525 Budapest, Hungary}
\date{\today}
\maketitle

\begin{abstract}
We investigate the influence of correlated initial conditions on
the temporal evolution of a ($d$+1)-dimensional critical directed 
percolation process. Generating initial states with correlations
$\langle s_i s_{i+r} \rangle \sim r^{\sigma-d}$ we observe  
that the density of active sites in Monte-Carlo simulations
evolves as $\rho(t) \sim t^{\kappa}$. The exponent $\kappa$ 
depends continuously on $\sigma$ and varies in the range
$-\beta/\nu_{||} \leq \kappa \leq \eta$. Our numerical results
are confirmed by an exact field-theoretical renormalization 
group calculation.
\end{abstract}

%
%
%
%
%
%
\section{Introduction}

It is well known that initial conditions influence the temporal
evolution of nonequilibrium systems. The systems' ``memory''
for the initial state usually depends on the dynamical rules.
For example, stochastic processes with a finite temporal correlation 
length relax to their stationary state in an exponentially short time.
An interesting situation emerges when a system undergoes a 
nonequilibrium phase transition where the temporal correlation
length diverges. This raises the question whether
it is possible construct initial states that affect the 
{\it entire} temporal evolution of such systems.

To address this question, we consider the example of
Directed Percolation (DP) which is the canonical universality class for
nonequilibrium phase transitions from an active phase into an absorbing
state~\cite{percol}. DP is used as a model describing the
spreading of some non-conserved agent and may be
interpreted as a time-dependent stochastic process 
in which particles produce offspring and self-annihilate.
Depending on the rates for offspring production and
self-annihilation such models display a continuous phase
transition from a fluctuating active phase into an absorbing state
without particles from where the system cannot escape.
Close to the phase transition the emerging critical behavior 
is characterized by a particle distribution with
fractal properties and long-range correlations. The DP
phase transition is extremely robust with respect to the microscopic
details of the dynamical rules~\cite{langevin,dpconj}
and takes place even in 1+1 dimensions.

Monte-Carlo (MC) simulations of critical models with absorbing states 
usually employ two different types of initial conditions.
On the one hand {\it random initial conditions} (Poisson
distributions)
are used to study the relaxation of an initial state with a finite 
particle density towards the absorbing state. 
In this case the particle density $\rho(t)$
{\it decreases} on the infinite lattice asymptotically as
(for the definition of the DP scaling
exponents $\beta$,$\nu_\perp$,$\nu_{||}$,$z$ see Ref.~\cite{percol})
\begin{equation}
\label{Decrease}
\rho(t) \sim t^{-\beta/\nu_{||}} \,.
\end{equation}
On the other hand, in so-called dynamic MC simulations~\cite{dynmcs},
each run starts with a {\it single particle} as a localized
active seed from where a cluster originates.
Although many of these clusters survive for only a short time, 
the number of particles $n(t)$ averaged over many
independent runs {\em increases} as
\begin{equation}
\label{Increase}
\langle n(t) \rangle \sim t^{+ \eta} \,,
\end{equation}
where $\eta = (\nu_\perp-2\beta)/\nu_{||}$.
These two cases seem to represent extremal situations where the
average particle number either decreases or increases.

A {\em crossover} between these two extremal cases takes place 
in a critical DP process that starts from a random initial condition
at very low density. Here the particles are initially 
separated by empty intervals of a certain typical size wherefore
the average particle number first increases 
according to Eq.~(\ref{Increase}). Later, when the growing 
clusters begin to interact with each other, the system crosses over
to the algebraic decay of Eq.~(\ref{Decrease}) -- a phenomenon which
is referred to as the ``critical initial slip'' of nonequilibrium
systems~\cite{crslip}. 

In the present work we investigate whether it is possible
to
interpolate {\em continuously} between the two extremal cases.
As will be shown, one can in fact generate 
certain initial states in a way
that the particle density on the infinite lattice varies as
\begin{equation}
\label{Decay}
\rho(t) \sim t^\kappa
\end{equation}
with a continuously adjustable exponent $\kappa$ in the range
\begin{equation}
\label{KappaRange}
-\beta/\nu_{||} \leq \kappa \leq +\eta \,.
\end{equation}
To this end we construct artificial initial configurations with
algebraic
long-range correlations of the form
\begin{equation}
\label{TwoPointCorrelations}
C(r) = \langle s_i \, s_{i+r} \rangle
\sim r^{-(d-\sigma)} \,,
\end{equation}
where $\langle\rangle$ denotes the average over many independent
realizations, $d$ the spatial dimension, and $s_i=0,1$ 
inactive and active sites. 
The exponent $\sigma$ is a free parameter and can be varied continuously
between $0$ and $1$. The limit of long-range correlations
$\sigma \rightarrow d$ corresponds to 
a constant particle density and thus we expect Eq.~(\ref{Decrease})
to hold. On the other hand, the short-range limit $\sigma \rightarrow 0$ 
represents an initial state where active sites are separated by 
infinitely large intervals so that the particle density should increase
according to Eq.~(\ref{Increase}). In between we expect 
$\rho(t)$ to vary algebraically according
to Eq.~(\ref{Decay}) with an exponent $\kappa$ 
depending continuously on $\sigma$. Our aim is to investigate the
functional dependence of $\kappa(\sigma)$.

The effect of power-law correlated initial conditions 
$\langle \phi (0) \phi (r) \rangle \sim r^{-(d-\sigma)}$ 
in case of a quench to the ordered phase of 
systems with nonconserved order parameter 
has been investigated some time ago by Bray et. al. \cite{bray}. 
Such systems are characterized by coarsening domains that
grow with time as $t^{1/2}$. An important example is the
(2+1)-dimensional Glauber-Ising model quenched to zero temperature.
It was observed that long-range correlations are relevant
only if $\sigma$ exceeds a critical value $\sigma_c$. Furthermore,
it was shown that the relevant regime is characterized by a continuously
changing exponent in the autocorrelation function
$A(t) = \left[ \phi (r,t) \phi (r,0) \right] \sim t^{-(d-\sigma)/4}$,
whereas the usual short-range scaling exponents 
could be recovered below the threshold. 
The results were found to be in agreement 
with the simulation results for the two-dimensional Ising model
quenched from $T=T_c$ to $T=0$.

The DP process -- the prototype of models with a phase 
transition from an active phase into an absorbing state -- 
is different from coarsening processes.
Instead of growing domains the DP process generates fractal
clusters of active sites with a coherence length $\xi_\perp$ 
which grows as $t^{1/z}$ where $z=\nu_{||}/\nu_\perp$. 
Thus the scaling forms assumed in
Ref.~\cite{bray} are no longer valid in the present case.
In addition, the field-theoretical description of DP involves
nontrivial loop corrections and thus we are interested
to find out to what extent the results are different from those
in Ref.~\cite{bray}. Our investigation also sheds some light
on the relation between the observed phenomena for correlated initial 
states, the critical initial slip, and scaling laws in
time dependent simulations.

In the present work we focus on the following aspects of the 
problem: In Sect.~\ref{InitCondSection} we describe in
detail how correlated initial states can be constructed 
in one dimension.
Using these states we then perform MC simulations in order to 
numerically estimate the exponent 
$\kappa$ as a function of $\sigma$ (see Sect.~\ref{NumSection}
and Fig.~\ref{kappa}). In Sect.~\ref{RGSection} we
present a field-theoretical renormalization 
group calculation which generalizes recent results 
obtained by Wijland et. al.~\cite{wijland}.
Because of a special property of the vertex diagrams
and the loop diagrams for the initial particle distribution
it is possible to derive an exact scaling relation,
leading to our main result
\begin{equation}
\label{ExactSolution}
\kappa(\sigma) = \left\{
\begin{array}{lll}
\eta & \mbox{for} & \sigma<\sigma_c \\[3mm]
\frac{1}{z} (d-\sigma-\beta/\nu_\perp) & \mbox{for} & \sigma>\sigma_c
\end{array}
\right.
\end{equation}
with the critical threshold $\sigma_c=\beta/\nu_\perp$. 
Because of the scaling relation $\eta = (\nu_\perp-2\beta)/\nu_{||}$
this function is continuous at $\sigma=\sigma_c$.
The theoretical result is found to be in agreement with our 
simulation results in one spatial dimension.
In Sect.~\ref{NaturalCorrSection} we compare the 
correlations of our constructed
initial states with the ``natural'' correlations that are
generated by the DP process itself. Finally we
summarize our conclusions in Sect.~\ref{ConclusionSection}.

\section{Construction of correlated initial states}
\label{InitCondSection}

The construction of artificial correlated particle
distributions on a lattice is a highly nontrivial task
since the lattice spacing and finite size effects lead
to deviations that strongly affect the accuracy of the
numerical simulations. Therefore one has to carefully
verify the correlation exponent and fractal dimension
of the generated distribution. In this Section we 
describe in detail how such particle distributions can be
generated and tested. For simplicity we restrict ourselves
to initial states in one spatial dimension.

Let us consider a particle distribution
on the real line where particles are separated by 
empty intervals of length $\ell$. We assume that these 
intervals are uncorrelated and power-law 
distributed according to 
\begin{equation}
\label{Distribution}
P(\ell) \sim \ell^{-\alpha}\,.
\qquad \qquad 
1 < \alpha \leq 2
\end{equation}
%
%
%
%
%
\begin{figure}
\epsfxsize=100mm
\centerline{\epsffile{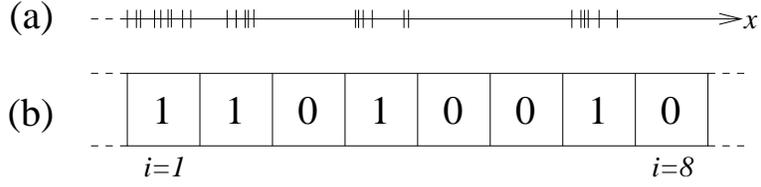}}
\caption{
Schematic illustration of the projection
from (a) an almost perfect fractal set 
onto (b) the lattice sites $s_1,\ldots,s_8$.
}
\label{FigProjection}
\end{figure}
%
%
%
%
%
%
\begin{figure}
\epsfxsize=150mm
\centerline{\epsffile{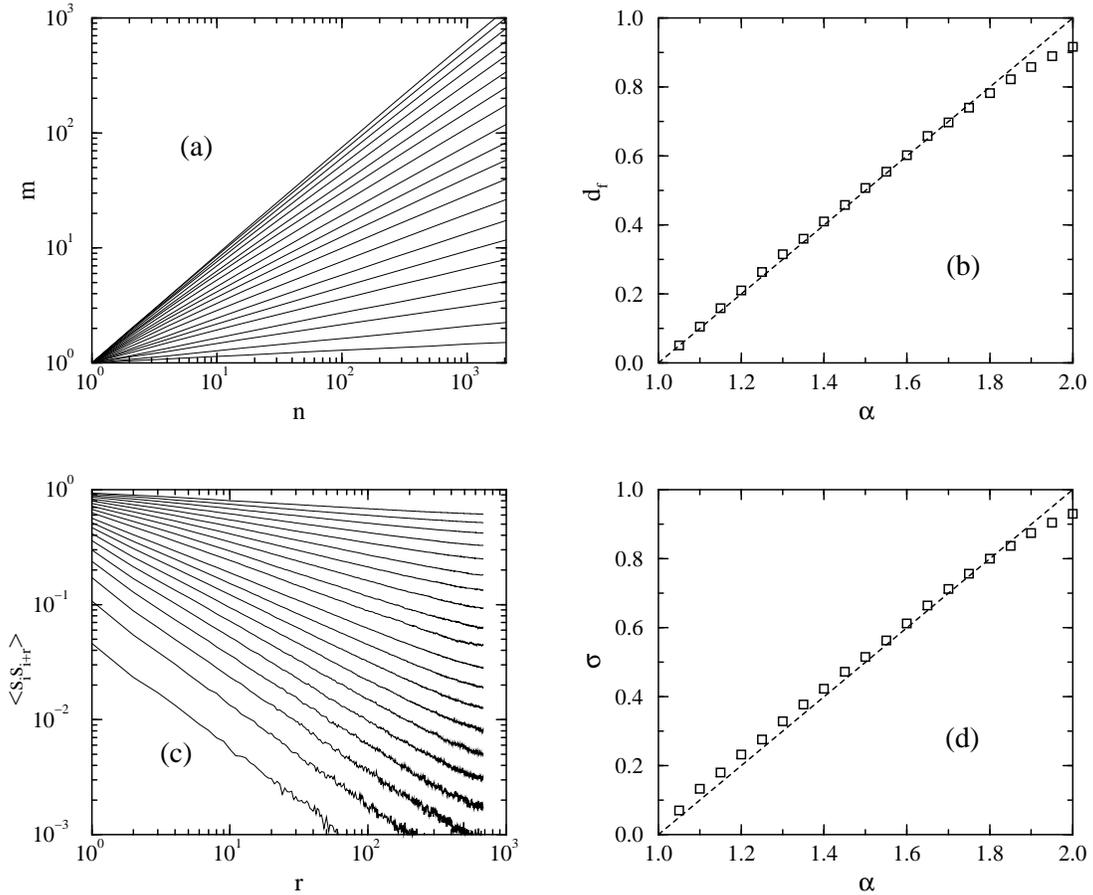}}
\caption{
Numerical verification of the fractal dimension~$d_f$ and
the correlation exponent~$\sigma$ on a lattice with $2^{11}$ sites
averaged over $20000$ samples.
(a): Number of active boxes~$m$ versus the total number of 
boxes~$n$ for various values of~$\alpha$. 
(b): Estimate of the corresponding fractal dimension~$d_f$.
(c)-(d): Analogous estimation of the correlation exponent~$\sigma$
as a function of~$\alpha$.
}
\label{FigFractaldim}
\end{figure}

This distribution corresponds to a simple 
fractal set with the fractal dimension $d_f=\alpha-1$, hence
the range of $\alpha$ is restricted by $1 < \alpha \leq 2$. 
On a lattice, however, the lattice spacing and 
the system size $L$ have to be taken into account as
lower and upper cutoffs for the distribution $P(\ell)$. 
The quality of a lattice approximation depends
on the actual implementation of these cutoffs.
It turns out that the accuracy of MC simulations depends
strongly on the quality of the initial states and therefore the
proper implementation of the cutoffs is crucial in the 
present problem. 

We find that a good approximation is obtained when an
(almost) perfect fractal set is projected onto the lattice
in a way that site $i$ becomes active if at least one element
of the fractal belongs to the interval $[i,i+1]$.
The resulting lattice configuration is the minimal
set of boxes on the lattice that is needed to
cover the fractal set.
This projection can be efficiently realized on a computer by
generating a sequence of points~$x$ on the real line 
separated by intervals distributed according to 
Eq.~(\ref{Distribution}) with a very small cutoff
$\Lambda_{min} \ll 1$ and projecting it onto the lattice by
the following prescription (see Fig.~\ref{FigProjection}):

\begin{enumerate}

\item
Start with the empty lattice $s_i=0$ (i=1,\ldots,L) and let
$x$ be a real variable with the initial value $x=1$. 

\item
Let $i$ be the maximal integer for which $i \leq x$
and turn site $i$ into the active state by setting $s_i=1$.

\item
Let $\Lambda_{max}=L-x$ be the current upper
cutoff and generate a random number $0<y<1$ from a flat distribution. If
$y<(\Lambda_{max}/\Lambda_{min})^{1-\alpha}$ the construction of
the initial state is finished, otherwise continue.

\item
Generate another random number $z$ from a flat distribution
in the interval $\Lambda_{max}^{1-\alpha}<z<\Lambda_{min}^{1-\alpha}$
and increment $x$ by $\ell=z^{1/(1-\alpha)}$, and continue at step $2$.

\end{enumerate}

Notice that step $3$ takes the upper cutoff into account by
finishing the loop when the generated interval $\ell$ would exceed the
remaining size of the chain $L-x$. The lower
cutoff is processed in step $4$ by truncating the allowed
range of $z$. 

In order to verify the quality of this approximation, we
numerically estimate the fractal dimension $d_f$ of the generated
initial states by box counting. To this end we divide the
whole system into $n$ boxes and count the number~$m$ of boxes 
that contain at least one active site, averaging over many
independent realizations. In Fig.~\ref{FigFractaldim}a $m$ is plotted
against $n$ in a double-logarithmic scale. The straight lines
indicate that the ``true'' fractal is well approximated. From 
the slopes we estimate the fractal dimension $d_f$ which
is shown in Fig.~\ref{FigFractaldim}b as a function of $\alpha$.
We also measure the two-point correlations in the generated states
which should be precisely those of 
Eq.~(\ref{TwoPointCorrelations}) with $\sigma=\alpha-1$. 
This can be proven by assuming that the intervals are
uncorrelated and evaluating a geometric series of the Laplace
transform of $P(\ell)$. In order to verify this relation, we estimated
$\sigma(\alpha)$ numerically in 
Fig.~\ref{FigFractaldim}c-\ref{FigFractaldim}d.

In both measurements we find a fairly good agreement with the exact 
results (dashed lines in Fig.~\ref{FigFractaldim}). It turns out
the deviations close to $\alpha=1$ can be reduced by increasing
the system size while the deviations close to $\alpha=2$ are
due to the lattice spacing and the lower cutoff $\Lambda_{min}$.

It should be emphasized that these artificial initial states have a
vanishing particle density in the limit $L \rightarrow \infty$.
On a finite lattice, however, a finite density is generated
which depends on $\alpha$ and may vary over several decades.
By increasing the lattice size we therefore reduce the initial 
particle density which leads to a higher statistical error in the
subsequent DP process. Thus the optimal system size has to be determined
by balancing discretization errors of the initial states against
statistical errors of the DP process.

\section{Numerical Results}
\label{NumSection}

The time dependent simulations have been performed by using
a Domany-Kinzel stochastic cellular automaton~\cite{DK}
with parallel updates and periodic boundary conditions.
The Domany-Kinzel model is controlled by two parameters
$p_1$ and $p_2$ and has a whole phase transition line
where a DP transition takes place. We performed simulations
for three different realizations, namely
\begin{itemize}
\item $p_1=0.8092$, $p_2=0$: \ \ Wolfram-18 transition,
\item $p_1=p_2=0.705485$; \ \ site percolation,
\item $p_1=0.644700$, $p_2=0.873762$: \ \  bond percolation.
\end{itemize}
The results turned out to be the same in all cases 
within numerical accuracy, although the bond and
site percolation showed a better scaling law behavior.
An efficient, multi-spin coded program has been employed that
simulates $32$ replicas for different values $\alpha\in (1,2)$, 
exploiting the advantage that the same random numbers 
can be used on each replica at a given site.
This allows us to update $32$ systems in parallel 
stored as an integer vector of length $L$. 
The lattice size has been varied between $L=128$ and $10000$,
while the number of independent samples was $10000 - 500$
respectively. Finite size effects turned out to be negligible
for $L \ge 4096$.
\begin{figure}
\epsfxsize=130mm
\centerline{\epsffile{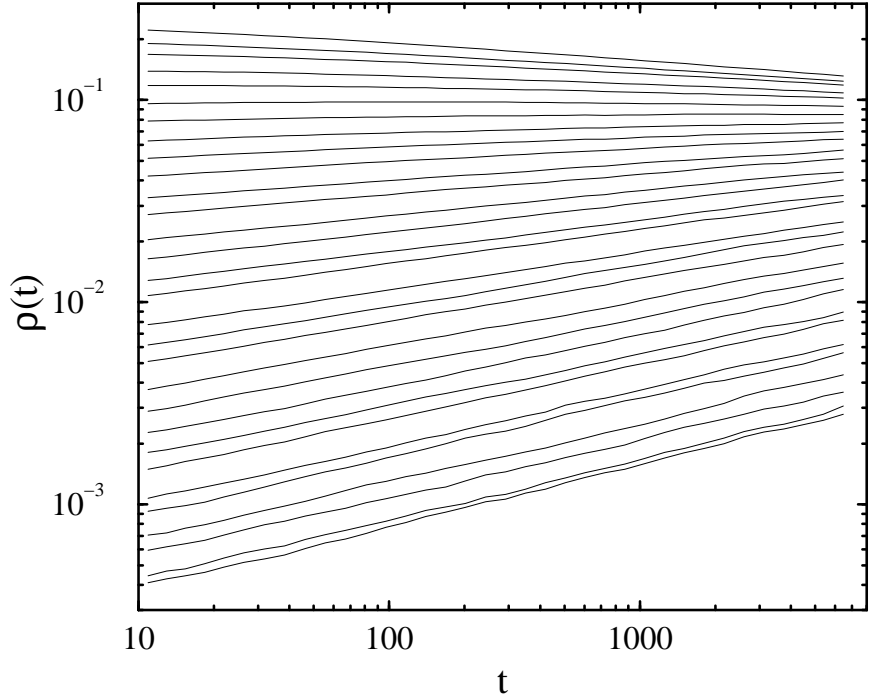}}
\caption{
Temporal dependence of $\rho$ for different
initial correlations characterized by $\alpha = 1...2$
in steps of $1/32$ (from bottom to top). 
The simulation has been performed
on a periodic lattice with $L=8192$ sites, 
a lower cutoff $\Lambda_{min}=0.0001$,
averaging statistically over $1600$ samples.
}
\label{rho}
\end{figure}
As shown in Fig.~\ref{rho} the quality of the power-law for
$\rho(t)$ is quite convincing and extends over the first three 
decades in the case of a system with $L=8192$ sites. 
For finite systems all these curves will eventually cross over 
to the $t^{-\beta/\nu_{||}}$ decay after very long time.
The slopes of the lines in the log-log plot -- measuring the exponent
$\kappa(\alpha)$ -- have been extrapolated from the 
last two decades by standard linear regression analysis. 
For each $\alpha$ we also determine the corresponding correlation
exponent $\sigma(\alpha)$ as explained in the previous Section
(see Fig.~\ref{FigFractaldim}). In Fig.~\ref{kappa} we plot
$\kappa(\alpha)$ versus $\sigma(\alpha)$. Using the numerically
known estimates for the DP exponents~\cite{Jensen}
\begin{equation}
\beta=0.2765\,,
\qquad
\nu_{||}=1.7338\,,
\qquad
\nu_\perp=1.0968\,,
\qquad
z=1.5807\,,
\qquad
\eta=0.3137\,,
\end{equation}
our results are in a fairly good agreement with the theoretical 
prediction of Eq.~(\ref{ExactSolution}) (solid line). 
The deviations for small $\sigma$ originate from the lattice cutoff
and could be reduced by further increasing the computational effort.
\begin{figure}
\epsfxsize=130mm
\centerline{\epsffile{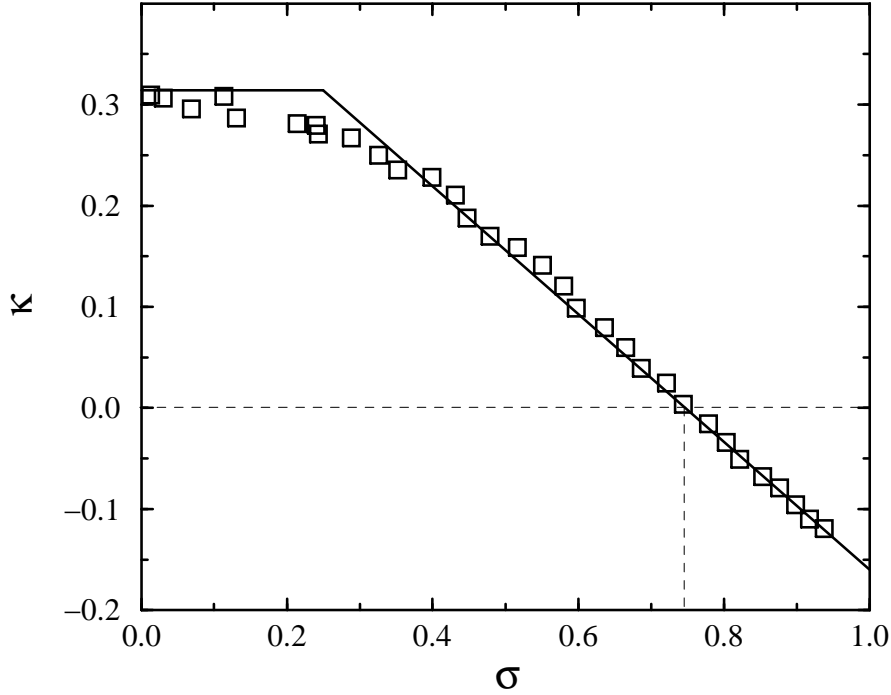}}
\caption{ 
Estimates for the exponents $\kappa(\sigma)$ obtained 
from the slopes of the lines in Fig.~\ref{rho}.
The theoretical prediction is shown by the solid line.
The dashed line indicates the ``natural'' correlations of DP
(see Sect.~\ref{ConclusionSection}).
}
\label{kappa}
\end{figure}
%
%
%
\section{Renormalization group calculation}
\label{RGSection}

In this Section we derive Eq.~(\ref{ExactSolution}) by an exact
RG calculation. To this end the DP Langevin equation~\cite{langevin}
has to be extended by an additional term 
\begin{equation}
\label{Langevin}
\partial_t \rho(x,t) = a \rho(x,t) - b \rho^2(x,t) + 
D \nabla^2 \rho(x,t)
+ \eta(x,t) + \Delta_1 \phi(x) \delta(t) \,,
\end{equation}
where $\phi(x)$ represents the initial particle distribution
and $\eta(x,t)$ is a $\rho$-dependent noise field with correlations
\begin{equation}
\langle \eta(x,t) \eta(x^\prime,t^\prime) \rangle
= \gamma \,\rho(x,t)\,\delta(x-x^\prime)\, \delta(t-t^\prime) \,.
\end{equation}
Although in principle we could directly analyze the Langevin equation,
it is often more convenient to derive the corresponding
effective field theory by introducing a second-quantized 
bosonic operator representation~\cite{effective}. The
Langevin equation it then transformed into an effective action
$S=S_0+S_{int}+S_{ipd}$, where $S_0$, $S_{int}$, and $S_{ipd}$
denote the free part, the non-linear interaction, and the
contribution for the initial particle distribution. Following 
the notation of Ref.~\cite{wijland}, the respective parts are given by:
\begin{eqnarray}
\label{BilinearContribution}
S_0[\psi,\bar{\psi}] &=& 
\int d^dx\,dt\,
\bar{\psi}(x,t)\, [ \partial_t + \lambda(\sigma-\nabla^2)] \, \psi(x,t)
\\
\label{Interaction}
S_{int} [\psi,\bar{\psi}] &=& 
\int d^dx\,dt\,
g \psi(x,t) \, \bar{\psi}(x,t) \, [\psi(x,t) - \bar{\psi}(x,t)]
\\
\label{InitialParticleDistribution}
S_{ipd} [\bar{\psi},\phi] &=& 
\int d^dx\,
\sum_{j=1}^\infty \Delta_j \bar{\psi}^j(x,0) \phi(x) \,.
\end{eqnarray}
The part $S_0+S_{int}$ is just the usual action of Reggeon field 
theory~\cite{regfth,langevin}. The additional contribution 
$S_{ipd}$ couples the field $\bar{\psi}(x,0)$ with
the initial particle distribution $\phi(x)$. It is written in its
most general form containing contributions of all orders
in $\bar{\psi}$ with independent coefficients $\Delta_j$. 
The lowest order contribution $\Delta_1\bar{\psi}\phi$ 
in Eq.~(\ref{InitialParticleDistribution})
corresponds to the term $\Delta_1 \phi(x) \delta(t)$
in the Langevin equation (\ref{Langevin}). The higher order terms
for $j \geq 2$ are included because they may be generated under
RG transformations. However, as we will see below, it is 
actually sufficient to consider the first two contributions.

The relaxation of a DP process with random
initial conditions $\phi(x)=\rho_0$
to its stationary state  has been studied recently by Wijland 
et. al.~\cite{wijland} using Wilson's dynamic renormalization 
group approach. At the critical dimension $d_c=4$ the constant
contribution $\Delta_1\bar{\psi}\phi$ was shown to be relevant while
the ``Poissonian'' contribution $\Delta_2 \bar{\psi}^2\phi$ turned out
to be marginal. For $d<d_c$, however, fluctuations corrections 
cause the coefficient $\Delta_2$ 
to vanish under RG transformations. This allowed the authors to 
express the so-called critical initial slip exponent $\eta$
($\theta'$ in their notation) by an exact scaling relation. 
As mentioned in the Introduction, this exponent 
describes the initial short-time behavior of $\rho(t)$ 
until the correlations generated by the dynamical process 
are longer than the typical size of empty intervals in the
initial state~\cite{crslip} so that the system crosses over
to the usual decay of Eq.~(\ref{Decrease}).
In the present case, however, the interval sizes of the initial 
state are power-law distributed and affect not only the 
initial but in principle the whole temporal evolution. 
Nevertheless we can use the formalism of Ref.~\cite{wijland} 
in order to determine the exponent $\kappa$ as a function of $\sigma$. 
The only difference is that we introduce a field $\phi(x)$ that
carries a non-trivial scaling dimension $d_\phi$.

Let us first consider a scaling transformation
\begin{equation}
x \rightarrow x^\prime = b x \,,
\qquad \qquad
t \rightarrow t^\prime = b^z t \,,
\end{equation}
where $z=\nu_\perp/\nu_{||}$ is the anisotropy exponent
of DP. Under this transformation 
the fields in Eqs.~(\ref{BilinearContribution})-
(\ref{InitialParticleDistribution}) change according to
\begin{eqnarray}
\label{rescaling}
\phi(x) &\rightarrow& \phi^\prime(x^\prime) = b^{-d_\phi} \phi(x)
\nonumber \\
\psi(x,t) &\rightarrow& \psi^\prime(x^\prime,t^\prime) = 
        b^{-d_\psi} \psi(x,t)  \\
\bar{\psi}(x,t) &\rightarrow& \bar{\psi}^\prime(x^\prime,t^\prime) = 
        b^{-d_{\bar{\psi}}} \bar{\psi}(x,t) \,. \nonumber
\end{eqnarray}
As in Reggeon field theory, the fields $\psi,\bar{\psi}$ carry the same
scaling dimension $d_\psi=d_{\bar{\psi}}$. 
The {\it scaling} dimension  of the initial state 
$d_\phi$ is related to the {\it fractal} 
dimension $d_f$  as follows. The number of active sites $N$
on a lattice with $L$ sites grows as $N \sim L^{d_f}$.
On the other hand, $N \sim \langle \phi \rangle L^d$,
where $\langle \phi \rangle$ denotes the average density of
active sites, i.e., $\langle \phi \rangle \sim L^{d_f-d}$.
Under rescaling this relation turns into $\langle \phi \rangle
b^{-d_\phi} = (Lb)^{d_f-d}$, hence
\begin{equation}
d_\phi = d-d_f = d-\sigma\,.
\end{equation}
Under rescaling the contribution $S_{ipd}$ changes by
\begin{equation}
S_{ipd} \rightarrow S^\prime_{ipd} = \int d^dx \,\sum_{j=1}^\infty \,
b^{d-jd_{\psi}-d_\phi} \, \Delta_j \, \bar{\psi}^j \phi(x) \,,
\end{equation}
i.e., the coefficients $\Delta_j$ transform according to
\begin{equation}
\Delta_j \rightarrow \Delta^\prime_j = 
b^{d-jd_{\psi}-d_\phi} \Delta_j \,.
\end{equation}
Defining the anomalous scaling dimensions $\eta_\psi, \eta_\phi$ of
the fields $\psi,\phi$ by
\begin{equation}
\label{AnomalousDimensions}
d_\psi = \frac12 (d+\eta_\psi)\,,
\qquad \qquad
d_\phi = \frac12 (d+\eta_\phi)
\end{equation}
the scaling dimensions of the coefficients $\Delta_j$
in the mean field approximation is given by
\begin{equation}
d_{\Delta_j}^{MF} = \frac12 \Bigl[d(1-j)-j\eta_\psi-\eta_\phi \Bigr]
\,.
\end{equation}
This result is expected to hold at the critical dimension $d_c=4$.
Notice that the relevance of the contributions $\Delta_j$ 
decreases with increasing~$j$. In systems with less than four 
spatial dimensions, fluctuation corrections have to be taken
into account. Assuming that contributions
with $j\geq3$ in Eq.~(\ref{InitialParticleDistribution}) 
are irrelevant, these corrections have been computed 
in Ref.~\cite{wijland} in a one-loop approximation. 
It was shown that in $d=4-\epsilon$ dimensions
the coefficients $\Delta_1$ and $\Delta_2$ change under 
infinitesimal scaling $b=1+l$ according to
\begin{eqnarray}
\label{Delta1}
\frac{d\Delta_1}{dl} \,&=&\,
\Delta_1\frac{-\eta_\psi-\eta_\phi}{2} \,+\,
\Delta_2 \frac{g K_4 \Lambda^2}{\lambda (\Lambda^2+\sigma)} \,-\,
\Delta_1\Delta_2 \frac{2 g^2 K_4}{\lambda^2} 
\,+\, 0(\epsilon^2)\\
\label{Delta2}
\frac{d\Delta_2}{dl} \,&=&\,
\Delta_2 \frac{-d-2\eta_\psi-\eta_\phi}{2} \,-\, 
\Delta_2 \frac{2 g^2 K_4}{\lambda^2} \,-\, 
(\Delta_2)^2 \frac{5 g^2 K_4}{\lambda^2}
\,+\, 0(\epsilon^2) \,,
\end{eqnarray}
where $K_4$ denotes the surface area of the unit sphere in four
dimensions divided by $(2\pi)^4$ and $\Lambda$ is the ultraviolet
cutoff in momentum space.

%
%
%
%
\begin{figure}
\epsfxsize=80mm
\centerline{\epsffile{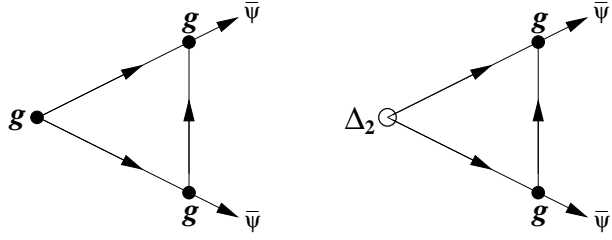}}
\vspace{3mm}
\caption{
One-loop diagrams for the nonlinear vertex $g$ (left)
and the Poissonian contribution $\Delta_2$ (right).
The diagrams are formally equivalent. The same applies to
higher loop diagrams that are linear in $\Delta_2$
and the corresponding vertex diagrams (not shown here).
Hence the loop corrections of these quantities are
are identical to all orders in $\epsilon$.
}
\label{FigDiagrams}
\end{figure}

Let us first consider the renormalization of $\Delta_2$.
As already noticed in Ref.~\cite{wijland}, the diagrams in 
Eq.~(\ref{Delta2}) which are linear in $\Delta_2$ are formally
identical with those for the renormalization of the 
nonlinear vertex $g$ to {\it all} orders in $\epsilon$ 
(see Fig.~\ref{FigDiagrams}). This observation plays a crucial role
in the present problem: Since the renormalization of the vertex
$g$ is given by
\begin{equation}
\label{VertexRenorm}
\frac{dg}{dl}=g\,\left[
z-\frac{d}{2}-\frac{3\eta_\psi}{2}-\frac{2g^2K_4}{\lambda^2}+\ldots
\right]
+ 0(\epsilon^2) 
\end{equation}
we may replace the diagrams of the form 
$\Delta_2 [-2 g^2 K_4 / \lambda^2 + 0(\epsilon^2)]$ 
in Eq.~(\ref{Delta2}) by the scaling part 
$\Delta_2 [z-d/2-3\eta_\psi/2]$ of Eq.~(\ref{VertexRenorm}), 
resulting in
\begin{equation}
\label{Delta2b}
\frac{d \Delta_2}{dl} \;=\;
\Delta_2 \left( \frac{\eta_\psi-\eta_\phi}{2} - z\right) +
(\Delta_2)^2 \Bigl[-\frac{5 g^2 K_4}{\lambda^2} + 0(\epsilon)\Bigr] \,.
\end{equation}
The linear part of this equation is now exact to all orders in
$\epsilon$. Since
\begin{equation}
\frac{\eta_\psi-\eta_\phi}{2} - z 
= d_\psi-d_\phi-z \leq d_\psi-z < 0
\end{equation}
we arrive at the conclusion that $\Delta_2$ scales to zero
for arbitrary initial states in $d<4$ spatial dimensions.
Thus Eq.~(\ref{Delta1}) reduces to its scaling part
and hence the scaling dimension of $\Delta_1$ is given by
\begin{equation}
d_{\Delta_1}=-\frac{\eta_\psi+\eta_\phi}{2}=
d-d_\psi-d_\phi = \sigma-d_\psi \,.
\end{equation}
Depending on the sign of $d_{\Delta_1}$, the loop corrections
for the initial particle distribution are {\it relevant}
for $\sigma>\sigma_c$, {\it irrelevant} for $\sigma<\sigma_c$,
and {\it marginal} for $\sigma=\sigma_c$.

Continuously changing exponents usually appear when we modify a model with
a marginal parameter which is invariant under RG transformations. For 
example, in case of models with infinitely many absorbing states the 
initial density plays that role~\cite{mendes}. In our case the two-point 
correlation of the  field $\psi(x,t)$ is characterized by $d_{\psi}=d-\sigma$ 
for $r > r_c(t)$ and is left invariant under the coarse-graining 
RG transformation, where $r_c(t) \to \infty$ 
is a growing spatial scale. In the infinite 
time limit this scaling crosses over to  $d_{\psi}=\beta/\nu_{\perp}$. 
The marginality of the operator $S_{ipd}$ requires
$\sigma-d_{\psi}=0$ and hence in the infinite time limit 
we arrive at
\begin{equation}
d_{\psi}(t\to\infty)=\sigma_c=\beta/\nu_{\perp}
\end{equation}
In the regime $\sigma>\sigma_c$ the exponent $\kappa$
is related to the scaling dimension $d_{\Delta_1}$
as follows. The particle density is expected to vary in the 
long time limit as
\begin{equation}
\label{rhooft}
\rho(t) \simeq \rho_0 \, t^{\kappa} \,,
\end{equation}
where $\rho_0 \propto \Delta_1$ is the 
initial particle density. Under
rescaling~(\ref{rescaling}) the density
$\rho(t)$ [which is essentially the average of $\psi(x,t)$]
scales as $\rho(t) \rightarrow b^{-d_\psi} \rho(t)$ while
the initial density transforms according to 
$\rho_0 \rightarrow b^{-d_\phi} \rho_0$. Thus scaling
invariance of Eq.~(\ref{rhooft}) requires that $-d_\psi=z\kappa-d_\phi$, 
i.e.,
\begin{equation}
\label{MarginalityCondition}
\kappa = \frac{d_\phi-d_\psi}{z} = \frac{\eta_\phi-\eta_\psi}{2z} \,.
\end{equation}
In the irrelevant regime $\sigma<\sigma_c$, however, the initial
density $\rho_0$ scales to zero under RG transformations which 
may be interpreted as an initial state with non-interacting 
active seeds leading to an increase according to Eq.~(\ref{Increase}).
Combining these results we arrive at Eq.~(\ref{ExactSolution})
which is exact to all orders in $\epsilon$.

\section{``Natural'' correlations in a DP process}
\label{NaturalCorrSection}

It is interesting to compare the artificial initial states of 
Sect.~\ref{InitCondSection} with ``natural'' DP scaling states. 
Scaling invariance predicts that
a critical DP process, starting from a fully occupied lattice, 
evolves towards a state with correlations
\begin{equation}
\label{NaturalCorrelations}
C(r) = \langle s_i \, s_{i+r} \rangle
\sim r^{-\beta/\nu_\perp}  \,, 
\qquad \qquad
r \ll \xi_\perp
\end{equation}
i.e., the ``natural'' correlations of the critical DP are given by
$\sigma^{DP} = d-\beta/\nu_\perp$. Interestingly,
the corresponding exponent $\kappa(\sigma^{DP})$ vanishes.
This means that these correlations characterize a situation
where the critical DP process is almost stationary. From 
the field-theoretical point of view this is not
surprising since $\kappa$ is determined solely by the
scaling dimension of the initial particle distribution.
Nevertheless the result is surprising from the physical point of
view since the empty intervals in our artificial initial states
are uncorrelated whereas such correlations may exist in DP scaling
states.
It seems that these correlations are rather weak~\cite{henkel} so that
states with uncorrelated intervals can be used as an approximation of
DP scaling states. This offers an interesting practical application
in DP simulations: Instead of starting a 
critical DP process from random initial conditions
and simulating over long transients, one could use the artificial
states of Sect.~\ref{InitCondSection} with $\sigma=\sigma^{DP}$ as
an approximation, followed by a short equilibration period to reach
the `true' scaling state of DP.

\section{Conclusions}
\label{ConclusionSection}

In the present work we numerically studied the temporal evolution
of a (1+1)-dimensional critical DP process starting from 
artificially generated correlated initial states. 
These states approximate a simple fractal set in which 
uncorrelated empty intervals are distributed by
$P(\ell) \sim \ell^{-\alpha}$. It can be shown
that such particle distributions are characterized
by long-range correlations of the form
$\langle s_is_{i+r} \rangle \sim r^{\sigma-1}$
where $\sigma=\alpha-1$. 

The construction of such correlated 
states is a technically difficult task
since lattice spacing and finite-size effects strongly
influence the quality of the numerical results, especially
close to $\alpha \approx 1$ and $\alpha \approx 2$.
In order to minimize these errors, we proposed to 
project an almost perfect fractal 
set onto the lattice. Using these states as 
initial conditions we determined the density~$\rho(t)$ in
a critical DP process by MC simulations. It varies algebraically
with an exponent $\kappa$ that depends continuously
on $\sigma$. Therefore correlated initial 
conditions affect the {\it entire} temporal evolution 
of a critical DP process.
The numerical estimates for $\kappa(\sigma)$ are
in good agreement with theoretical results
from a field-theoretical RG calculation (see Fig.~\ref{kappa}). 
Discrepancies for small values of $\sigma$
can be further reduced by increasing the numerical effort. 

The RG calculation is valid for arbitrary spatial dimensions $d<4$
and predicts a critical threshold 
$\sigma_c=\beta/\nu_\perp$ where $\kappa(\sigma)$ starts
to vary {\it linearly} between $+\eta$ and $-\beta/\nu_{||}$.
Thus our result in Eq.~(\ref{ExactSolution}) 
qualitatively reproduces the scenario predicted by Bray 
et.~al. in the context of coarsening processes~\cite{bray}. 
The exact prediction the critical threshold $\sigma_c$ 
is possible because of the irrelevance of $\Delta_2$ for
arbitrary initial conditions which
can be proven by using the formal equivalence 
of the loop expansions in $g$ and $\Delta_2$. It
is this property that also allows one to express the 
critical initial slip exponent $\eta$ by an 
exact scaling relation $\eta=(\nu_\perp-2\beta)/\nu_{||}$.
It should be emphasized that some DP models with
more complicated dynamical rules violate this scaling relation
as, for example, the two-species spreading model discussed in 
Ref.~\cite{wijland}. In such models the mentioned
equivalence of loop expansions in Fig.~\ref{FigDiagrams}
is no longer valid. Consequently, 
the corresponding slip exponent $\eta$ takes a different value
so that Eq.~(\ref{ExactSolution}) has to be modified appropriately.

\vspace{3mm}
Acknowledgements:
We would like to thank M. Howard, K. B. Lauritsen, N. Menyh\'ard, 
I. Peschel, and U. C. T\"auber for 
interesting discussions and helpful hints. 
The simulations were performed partially on the FUJITSU AP-1000
parallel supercomputer. G. \'Odor gratefully acknowledges
support from the Hungarian research fund OTKA ( Nos. T025286
and T023552) .


\end{document}